# MOLECULAR OXYGEN IN OORT CLOUD COMET 1P/HALLEY

M. Rubin[1], K. Altwegg[1,2], E. F. van Dishoeck[3], G. Schwehm[4]

[1] Physikalisches Institut, University of Bern, Sidlerstrasse 5, CH-3012 Bern, Switzerland
[2] Center for Space and Habitability, University of Bern, Sidlerstrasse 5, CH-3012 Bern, Switzerland
[3] Leiden Observatory, Leiden University, P.O. Box 9513, 2300 RA Leiden, The Netherlands
[4] ESA (retired) Science Operations Department, ESTEC, Keplerlaan 1, 2201 AZ Noordwijk, The Netherlands

## ABSTRACT

Recently the ROSINA mass spectrometer suite on board the European Space Agency's Rosetta spacecraft discovered an abundant amount of molecular oxygen, $O_2$, in the coma of Jupiter family comet 67P/Churyumov-Gerasimenko of $O_2/H_2O$ = 3.80±0.85%. It could be shown that $O_2$ is indeed a parent species and that the derived abundances point to a primordial origin.

One crucial question is whether the $O_2$ abundance is peculiar to comet 67P/Churyumov-Gerasimenko or Jupiter family comets in general or whether also Oort cloud comets such as comet 1P/Halley contain similar amounts of molecular oxygen. We investigated mass spectra obtained by the Neutral Mass Spectrometer instrument obtained during the flyby by the European Space Agency's Giotto probe at comet 1P/Halley. Our investigation indicates that a production rate of $O_2$ of 3.7±1.7% with respect to water is indeed compatible with the obtained Halley data and therefore that $O_2$ might be a rather common and abundant parent species.

## 1. INTRODUCTION

Bieler et al. (2015) showed that molecular oxygen represents the 4$^{th}$ most abundant species on average in the coma of comet 67P/Churyumov-Gerasimenko (hereafter 67P), after the well-known major species water, carbon monoxide, and carbon dioxide. Molecular oxygen has not been previously identified as a parent species in comets, possibly due to the difficulty in the detection using Earth-based telescopes as $O_2$ is abundant in Earth's atmosphere and lacks a strong dipole and therefore prominent rotational lines. Furthermore, excited oxygen atoms, often used as tracer, can form through photodissociation of not only $O_2$ but also other oxygen-containing species such as the major components in the coma $H_2O$, CO, and $CO_2$ (Huebner & Mukherjee 2015). In situ mass spectrometry at comet 1P/Halley lacked the required mass resolution and therefore contributions to the signal on the mass/charge 32 u/e channel could not be separated according to sulfur, $^{32}S$, methanol, $CH_3OH$, or molecular oxygen.

The amount of molecular oxygen found in 67P is rather surprising as it exceeds current Solar System formation model expectations (Zheng et al. 2006).

Furthermore the low relative abundances of the hydroperoxyl radical, $HO_2/O_2$ = $(1.9\pm0.3)\times10^{-3}$, hydrogen peroxide, $H_2O_2/O_2$ = $(0.60\pm0.07)\times10^{-3}$, as well as the lack of ozone, $O_3$, seem to be in contradiction with at least some of the expectations derived from numerical models and radiation experiments in the laboratory. UV irradiation and radiolysis of ices with subsequent sputtering is the major production process of the $O_2$ dominated atmosphere of the Jovian moon Europa (Hall et al. 1995). The same process is also at work at other icy moons around Jupiter and Saturn, such as Ganymede, Callisto, Dione, and Rhea driven by energetic particles in the Jovian as well as the Saturnian magnetospheres, respectively (e.g. Spencer et al. (1995); Noll et al. (1997); Sieger et al. (1998)).

Comets, however, spend most of their time in much less harsh environments: the irradiation by solar wind protons and alpha particles is less energetic, which results in very low penetration depths (Johnson et al. 1983). In addition, while ice has been discovered in some locations on the surface (Pommerol et al. 2015) most of it is covered under a crust (Capaccioni et al. 2015) and therefore not accessible to the low penetrating irradiation.

Also, during each apparition comets loose part of their mass through sublimation of volatiles which drag with them dust particles of various sizes. In the case of comet 67P, the effect of erosion ranges from meters to tens of meters (Keller et al. 2015) per apparition. 67P has a current perihelion distance of roughly 1.25 AU after a close encounter with Jupiter in 1959 (Carusi et al. 1985). With an orbital period of ~6.55 years this leads to almost 10 apparitions in its present orbit during which the comet lost tens of meters of surface material. This also poses a problem for explanations invoking the more energetic Galactic Cosmic Rays as their accumulated fluxes over the course of 67P's orbital period are too low for an efficient build-up of the observed amounts of $O_2$. Furthermore, the comet spent at least another 250 years, possibly up to 5000 years (Maquet 2015) at an intermediate perihelion distance inside Jupiter's orbit, therefore an outer layer of water ice with abundant amounts of $O_2$ is lost by now.

Comet 1P/Halley, however, has visited the inner Solar System much more often despite an orbital period more than ten times longer (75.3 years). Hughes (1985) estimated some 2000 apparitions in its current orbit (see also Olsson-Steel (1988) and references therein): combined with its perihelion distance of 0.59 AU comet 1P/Halley shows much higher gas production rates (Krankowsky et al. 1986) and it can be expected that during each apparition new material is exposed to sublimation. One possible explanation is that the $O_2$ has already been formed through irradiation of ices in the molecular cloud phase and the $O_2$ remained trapped (d'Hendecourt et al. 1985) before the comet eventually formed, although more recent astrochemical models (Taquet et al. 2012) cannot reproduce quantitatively the high observed $O_2/H_2O$ ratios in 67P/Churyumov-Gerasimenko by Bieler et al. (2015). Zheng et al. (2006) irradiated crystalline ices with electrons and derived $O_2/H_2O$ abundances of 0.6%, thus also falling short. Moreover the laboratory amounts of $H_2O_2$ and $HO_2$ with respect to $O_2$ are much higher compared to the observed ratios at comet 67P. Given the difficulty in explaining the high $O_2$ abundances observed in comet 67P, it is important to verify whether this comet is an exception or whether high levels of $O_2$ are common in comets.

## 2. MASS SPECTROMETRY DURING GIOTTO MISSION TO COMET 1P/HALLEY

On 14 March 1986 shortly past midnight the European Space Agency's Giotto mission flew past comet 1P/Halley with a closest approach of roughly 600 km (Reinhard 1986). This is roughly 1 month after 1P/Halley passed perihelion on 9 February 1986. The large relative velocity of 68 km/s was responsible for several payload instruments being rendered inoperable due to high velocity grain impacts. Still there is abundant data of the inbound path available from the Giotto NMS (Krankowsky et al. 1986). The NMS Mass-analyzer (M-analyzer) contained a fly-through electron impact ion source and a double focusing magnetic analyzer section followed by a focal plane detector. The mass range extends from 1 – 37 u/e with a FWHM mass resolution of $\Delta m \sim 0.3$ u/e.

The NMS Energy-analyzer (E-analyzer) consisted of a parallel plate energy analyzer coupled with a focal plane detector. Given the high velocity of the Giotto spacecraft with respect to the neutral gas particles the energy analysis is essentially equivalent to a mass analysis. The mass range extends from 8 – 86 u/e for neutrals and 1 – 56 u/e for ions.

Both analyzers used either 17 or 90 eV electron impact energy in neutral mode. The NMS sensor could also be operated in ion mode for which only charged particles from the coma could be detected. The NMS sensor obtained both, ion and neutral mass spectra, during the fly-by and the combination of both datasets allows for investigating the detailed chemical reactions occurring in the coma of an active comet such as 1P/Halley.

The limited mass resolution of NMS did not allow for distinguishing atoms and molecules of close molecular mass, such as sulfur, molecular oxygen, and methanol, all on mass/charge 32 u/e. The neutral gas production rates of methanol and hydrogen sulfide, $H_2S$, the latter a parent species of atomic sulfur, have therefore both been estimated by the amounts of their protonated counterparts $CH_3OH_2^+$ on mass/charge 33 u/e and $H_3S^+$ on mass/charge 35 u/e by means of an ion – neutral chemical network. This was then compared to NMS measurements in ion mode, which detects only particles already ionized in the coma (Eberhardt et al. 1994). From this comparison, relative production rates with respect to water of 1.71±0.04% and 0.41±0.02% for $CH_3OH$ and $H_2S$ have been derived, respectively. Both ion profiles are essentially compatible with $CH_3OH$ and $H_2S$ originating from the nucleus only. However, when again analyzing NMS data obtained in neutral mode, i.e. only neutral particles from the coma are detected, it became clear that the mass/charge 32 u/e channel still lacks a significant contribution (Rubin et al. 2011). From what we know now from comet 67P, the likely missing component on mass/charge 32 u/e at 1P/Halley is molecular oxygen.

## 3. DATA ANALYSIS

During the NMS laboratory calibration campaign all species, independent of their mass, were calibrated relative to molecular nitrogen ions, i.e. $N_2^+$ from ionization of $N_2$ (Meier 1992). This was necessary because it is impossible to distinguish species of similar mass with NMS, e.g. molecular oxygen and methanol. All species have different detection efficiencies and ionization cross sections, i.e. the same absolute densities of methanol and molecular oxygen lead to different signals on the NMS MCP detector.

As pointed out above, the volatile composition cannot be derived unambiguously with NMS. One can, however, guess the composition, e.g. from a numerical model, and compare to the NMS observations by applying the corresponding calibration factors and adding up the individual contributions to a single mass channel (e.g. sulfur, molecular oxygen, and methanol on mass/charge 32 u/e). Table 1 lists the detection efficiencies of the species used in this work relative to $N_2$ and $N_2^+$, respectively. Table 2 lists the differential impact ionization cross-sections for 90 eV electrons used in the ionization process. An example is discussed in the following paragraph.

When NMS is operated in neutral mode, first the fragmentation of the parent molecule inside the ion source has to be considered. For example, on mass/charge 32 u/e, less than 25% of the ionized $CH_3OH$ ends up as $CH_3OH^+$, the rest goes into $H_3CO^+$ on mass/charge 31 u/e, to $HCO^+$ on mass/charge 29 u/e, and so on (Pal 2004). The ionization cross section of 1.16 Å$^2$ for $CH_3OH \rightarrow CH_3OH^+$ is smaller than the 2.09 Å$^2$ for $N_2 \rightarrow N_2^+$ (Krishnakumar & Srivastava 1990), the species used for the calibration of NMS (Table 2). Also the detection efficiency of $CH_3OH^+$ is slightly smaller than for $N_2^+$ (97%, Table 1). The relative signal strength of $CH_3OH$ with respect to $N_2$ is therefore 0.54, meaning that a larger amount of methanol is needed for the same signal strength as $N_2$ (or $O_2$ for that matter which has to be corrected by 0.74 with respect to $N_2$). This way, by normalization of an assumed composition of multiple species to $N_2$, a synthetic signal can be derived for comparison to the obtained NMS measurements.

Such an analysis has been performed by Rubin et al. (2011) by use of a Direct Simulation Monte Carlo approach. Here we use the modeled gas velocities from this earlier work combined with a Haser model (Haser 1957) and complement the considered species by molecular oxygen, $O_2$, and the two sulfur bearing parent species hydrogen sulfide, $H_2S$, and carbonyl sulfide, $CS_2$, and their daughter species. We furthermore take the sulfur and oxygen isotopes into account, for which we assume solar abundances (sulfur: Altwegg (1995); $^{18}O/^{16}O$: Balsiger et al. (1995); Eberhardt et al. (1995)) and for the D/H ratio in methanol we use $3\times10^{-4}$ (Balsiger et al. 1995; Eberhardt et al. 1995) assuming a similarly elevated ratio as in the water of comet 1P/Halley. Note that the results are quite insensitive to the D/H in methanol and do not exclude D/H ratios on the order of $10^{-2}$ observed in interstellar ices (Charnley et al. 1997).

Table 3 lists the abundances of the species considered in this work relative to water. The corresponding mass/charge 18 u/e profile, which is dominated by cometary water, is shown in the left plot in Figure 1. The dotted blue line shows the modeled water density and the red solid line denotes the corresponding synthetic NMS signal obtained by the normalization to $N_2$ discussed above. Both lines are close as both the detector yields for $N_2^+$ and $H_2O^+$ (Table 1) and the cross sections for electron impact ionization inside the ion source for $N_2 \rightarrow N_2^+$ and $H_2O \rightarrow H_2O^+$ (Table 2) are quite close. The red solid line can be compared to NMS measurements from both the M-analyzer and E-analyzer represented by the red points (see also Krankowsky et al. (1986)). The plot on the right shows the same for the mass/charge 32 u/e profile. The modeled neutral parent densities of molecular oxygen and methanol are shown together with the synthetic signal to be compared to the NMS measurements. The red solid line is the modeled synthetic NMS signal and contains all contributions including methanol, molecular oxygen, and sulfur and the red zone includes the reported uncertainty of $O_2/H_2O$ = 3.7±1.7%. The red dashed line contains only methanol and parent species producing sulfur, $^{32}S$, and the red dotted line considers only methanol for comparison to Rubin et al. (2011).

Figure 2 shows both mass/charge channels 33 u/e and 34 u/e, respectively. Both figures have been used to derive upper limits for $H_2S$, which produces fragments on both channels (Table 2), and furthermore consider the contributions of the major isotopologues of methanol, molecular oxygen, and hydrogen sulfide. Molecules not considered here are $H_2O_2$ and $HO_2$ for which low abundances were derived at comet 67/Churyumov-Gerasimenko. At comet 1P/Halley the abundance could be higher; however, the limited mass resolution of NMS prevents an unambiguous determination.

The synthetic curves in Figure 1 show that methanol alone clearly underestimates the measured signal; in fact, if only methanol is considered, an abundance of approximately 7.2% with respect to water would be required, especially due to the smaller electron impact ionization cross section of the channel $CH_3OH \rightarrow CH_3OH^+$ compared to $O_2 \rightarrow O_2^+$ (Table 2). The addition of sulfur is of minor importance: the contribution of 1.2% $H_2S$ to the NMS signal corresponds to the difference between the red dotted line (methanol only) and the red dashed line (methanol plus sulfur from hydrogen sulfide and carbonyl sulfide) in Figure 1 (right). The $O_2/H_2O$ ratio is therefore almost independent whether the $H_2S$ production rate by Eberhardt et al. (1994) is used ($H_2S/H_2O$ = 0.41%) or our upper limit ($H_2S/H_2O$ = 1.2%). In summary, the combined analysis of 32, 33, and 34 u/e strengthens the conclusion that $CH_3OH$ and S are not sufficient to explain the 32 u/e signal.

## 4. DISCUSSION AND CONCLUSIONS

Neutral mass spectrometer data obtained during the Giotto fly-by are consistent with abundant amounts of $O_2$ in the coma of comet 1P/Halley. The inferred ratios are similar to the abundances observed for comet 67P/Churyumov-Gerasimenko: Bieler et al. (2015) derived a relative abundance of $O_2/H_2O$ = 3.80±0.85% which compares well to $O_2/H_2O$ = 3.7±1.7% for 1P/Halley. For Halley this makes $O_2$ the third most abundant species (Eberhardt 1999; Rubin et al. 2011). Note that the Halley data have been derived during the Giotto flyby lasting only a few minutes. Assuming a gas expansion velocity of approximately 1 km/s and the cometocentric distance for which we have data for the mass/charge 32 u/e channel (1,000 – 14,000 km) this corresponds to volatiles released from the comet during a period of ~4 hours. Our $O_2/H_2O$ value and associated error is therefore only a snapshot in time. The Rosetta results, on the other hand, are based on data acquired over several months at varying heliocentric distances and locations of the spacecraft with respect to the comet.

We now have an indication for abundant $O_2$ in the comae of two comets, one from the Oort cloud and the other from the Kuiper belt or possibly the scattered disk. This is particularly interesting, as both families of comets are believed to have formed at different locations in our early Solar System, supported by the difference in their D/H ratio in water (cf. Balsiger et al. (1995); Eberhardt et al. (1995); Altwegg et al. (2015)). This furthermore suggests that $O_2$ might be present in other comets and that perhaps 67/Churyumov-Gerasimenko and 1P/Halley are not unique. More generally, if icy planetesimals commonly contain $O_2$ at the level of a few percent, this $O_2$ may then also be delivered to planetary atmospheres that are built from such planetesimals. This in turn implies that not all atmospheric $O_2$ is necessarily a sign of biological activity.

The close abundance despite very different dynamical histories and erosion rates of both comets indicates that the observed $O_2$ has already been formed in the ices of the pre- and protosolar nebula, before the comet ultimately formed. This would in turn require that ice grains did not, or only partially, sublimate and reform during the collapse of the protosolar nebula, perhaps due to formation of planetesimals at a very early stage. Other scenarios are discussed in Bieler et al. (2015), who favor a relatively warm formation in the 20-30 K range (Rubin et al. 2015). Further evidence is required to test these scenarios, especially by the Rosetta mission at comet 67P/Churyumov-Gerasimenko in the coming year with fresh ice exposed following erosion during the August 2015 perihelion passage.

**Table 1:** Detection efficiencies of various species impinging on the NMS detector relative to molecular nitrogen ions, $N_2^+$ (Meier 1992).

| Species | Mass/charge | Detection efficiency |
|---|---|---|
| $H_2O^+$ | 18 u/e | 0.92 |
| $N_2^+$ | 28 u/e | 1.00 |
| $CH_3OH^+$ | 32 u/e | 0.97 |
| $O_2^+$ | 32 u/e | 1.00 |
| $S^+$ | 32 u/e | 0.73 |
| $HS^+$ | 33 u/e | 0.78 |
| $H_2S^+$ | 34 u/e | 0.81 |

**Table 2:** Electron impact ionization cross sections of neutrals inside NMS' ion source for the formation of fragment ions.

| Neutral parent | Fragment ion | Cross section [Å$^2$] | Reference |
|---|---|---|---|
| $N_2$ | $N_2^+$ | 2.090 | Krishnakumar & Srivastava (1990) |
| | $N_2^{++}$+$N^+$ | 0.623 | Krishnakumar & Srivastava (1990) |
| | $N^{++}$ | 0.005 | Krishnakumar & Srivastava (1990) |
| $CH_3OH$ | $CH_3OH^+$ | 1.164 | Pal (2004) |
| | $H_3CO^+$ | 1.644 | Pal (2004) |
| | $H_2CO^+$ | 0.125 | Pal (2004) |
| | $HCO^+$ | 0.940 | Pal (2004) |
| | $CO^+$ | 0.022 | Pal (2004) |
| | $OH^+$ | 0.028 | Pal (2004) |
| | $CH_3^+$ | 0.674 | Pal (2004) |
| | $CH_2^+$ | 0.079 | Pal (2004) |
| | $CH^+$ | 0.018 | Pal (2004) |
| | $C^+$ | 0.008 | Pal (2004) |
| | $H_2^+$ | 0.011 | Pal (2004) |
| | $H^+$ | 0.087 | Pal (2004) |
| $O_2$ | $O_2^+$ | 1.550 | Itikawa (2009) |
| | $O^+$ | 0.827 | Itikawa (2009) |
| | $O^{++}$ | 0.004 | Itikawa (2009) |
| $H_2S$ | $H_2S^+$ | 1.932 | Rao & Srivastava (1993) |
| | $HS^+$ | 0.824 | Rao & Srivastava (1993) |
| | $S^+$ | 0.872 | Rao & Srivastava (1993) |
| | $H^+$ | 0.169 | Rao & Srivastava (1993) |
| | $H_2^+$ | 0.008 | Rao & Srivastava (1993) |
| | $H_2S^{++}$ | 0.026 | Rao & Srivastava (1993) |
| | $S^{++}$ | 0.010 | Rao & Srivastava (1993) |
| HS | $HS^+$ | 2.000 | estimate |
| | $S^+$ | 1.000 | estimate |
| $CS_2$ | $CS_2^+$ | 4.480 | Lindsay et al. (2003) |
| | $S_2^+$ | 0.078 | Lindsay et al. (2003) |
| | $CS^+$ | 1.430 | Lindsay et al. (2003) |
| | $S^+$ | 2.305 | Lindsay et al. (2003) |
| | $C^+$ | 0.313 | Lindsay et al. (2003) |
| | $CS_2^{++}$ | 0.179 | Lindsay et al. (2003) |
| $H_2O$ | $H_2O^+$ | 2.115 | Rao et al. (1995) |
| | $OH^+$ | 0.610 | Rao et al. (1995) |
| | $O^+$ | 0.125 | Rao et al. (1995) |
| | $H^+$ | 0.267 | Rao et al. (1995) |

**Table 3:** Relative abundances of the parent species for 1P/Halley considered in this work and 67P/Churyumov-Gerasimenko for comparison.

| Species | Relative abundance 1P/Halley | Relative abundance 67P/Churyumov-Gerasimenko |
|---|---|---|
| $H_2O$ | 100% [a)] | 100% [a)] |
| $O_2$ | 3.7±1.7% [b)] | 3.80±0.85% [c)] |
| $CH_3OH$ | 1.71±0.04% [d)] | 0.31-0.55% [e)] |
| $H_2S$ [f)] | 0.41±0.02% [d)] | 0.67-1.75% [e)] |
| | 1.2% [g)] | |
| $CS_2$ | 0.2±0.1% [h)] | 0.003-0.024% [e)] |

[a] definition
[b] this work
[c] Bieler et al. (2015)
[d] Eberhardt et al. (1994)
[e] Observed ranges from Le Roy et al. (2015) above the summer and winter hemispheres
[f] For $H_2S$ we investigated both the production rate reported by Eberhardt et al. (1994) and an upper limit assuming that $H_2S$ dominates most of the mass/charge = 34 u/e channel
[g] upper limit, this work
[h] Feldman et al. (1987)

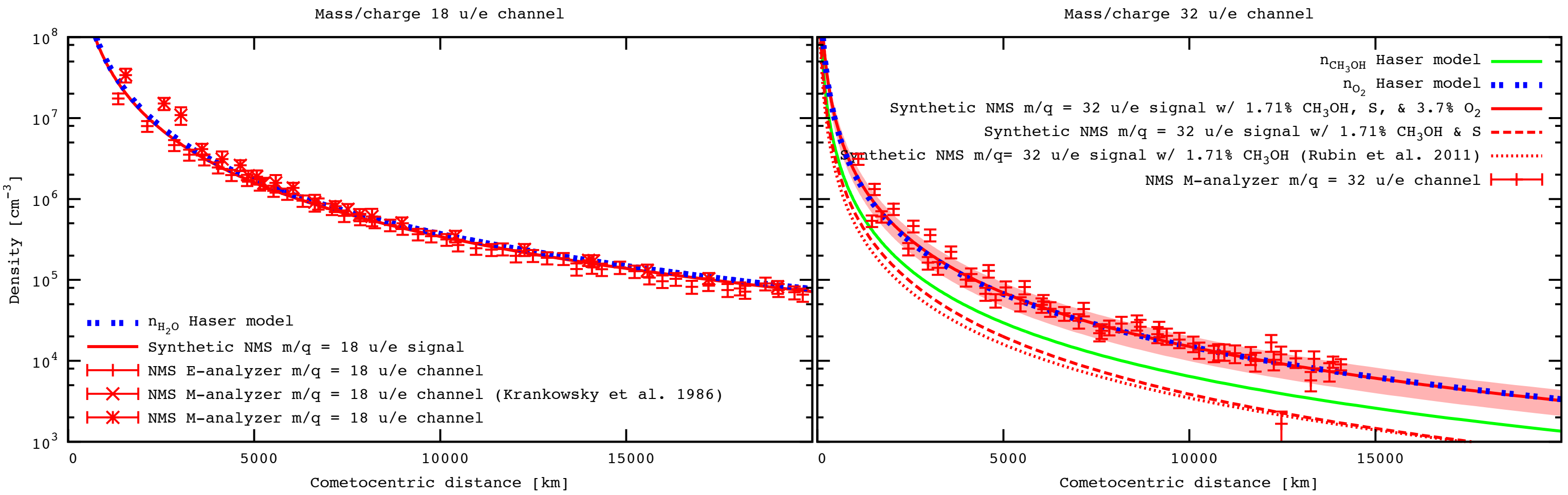

**Figure 1:** On the left is the mass per charge 18 u/e channel. The blue dotted line shows the Haser model water density while the red solid line shows the corresponding synthetic signal to be compared to the actual Giotto NMS measurements (red points). The mass per charge 32 u/e channel on the right shows the Haser model densities of methanol (solid green), molecular oxygen (dotted blue), and the corresponding synthetic signal including all contributions including isotopes (red solid), only methanol and sulfur (red dashed), and limited to methanol (red dotted). The uncertainty/variation of the $O_2/H_2O$ ratio of 3.7±1.7% is indicated by the red zone around the synthetic signal for comparison with the Giotto NMS signal (red points).

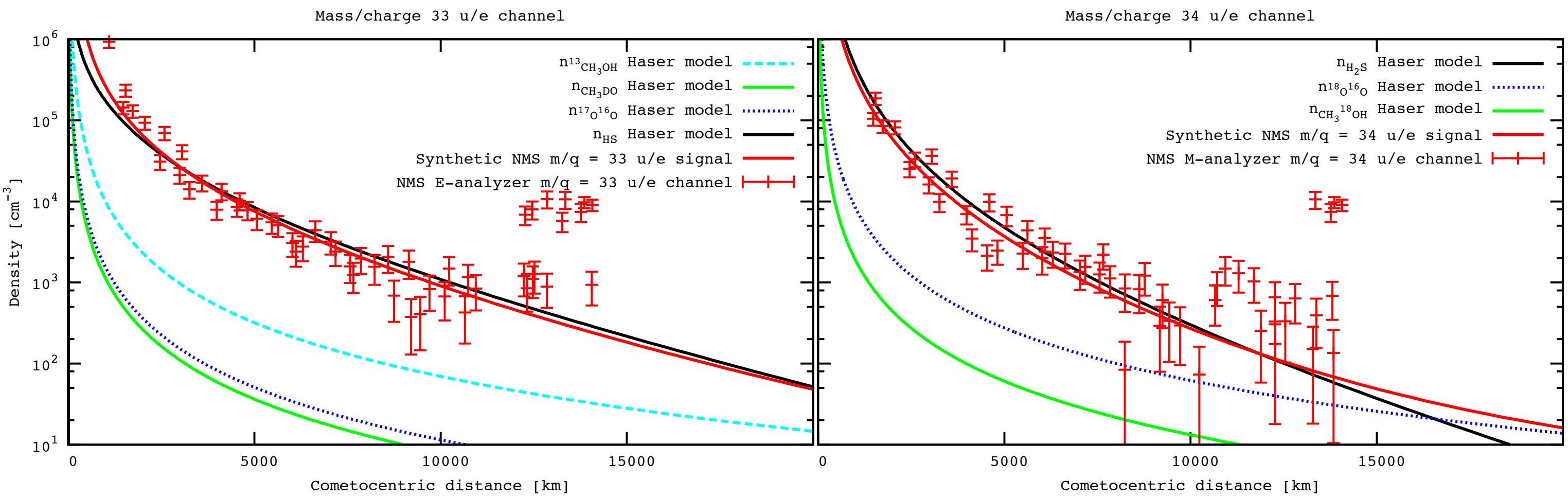

**Figure 2:** Left: Mass per charge 33 u/e channel with the main signal HS, fragment of $H_2S$. Right: Mass per charge 34 u/e channel with the main signal $H_2S$. For comparison also methanol and molecular oxygen are given when taking isotopic ratios into account. The red solid lines in both plates are the combined synthetic signal for comparison to the Giotto NMS measurements (red points). Note that NMS cannot distinguish between the possible molecular structures of deuterated methanol.